\documentclass[%
 reprint,
 amsmath,amssymb,
prx,
floatfix,
]{revtex4-2}

\usepackage{mathtools}
\usepackage{graphicx}
\usepackage{dcolumn}
\usepackage{bm}

\usepackage{qcircuit}
\usepackage[caption=false]{subfig}

\usepackage{algorithm}
\usepackage[noend]{algpseudocode}
\algrenewcommand\algorithmicdo{}

\makeatletter
\renewcommand{\ALG@name}{Procedure}
\makeatother

\usepackage[linktocpage=true,
  colorlinks=true, 
  pdfborder={0 0 0},
  linkcolor=blue,
  citecolor=red,
  filecolor=yellow,
  urlcolor=blue,
  bookmarks,
  pdfauthor={},
]{hyperref}

\usepackage{etoolbox}
\newcounter{is_pdf_used_for_fig}
\begin{document}

\preprint{APS/123-QED}

\title{
Imaginary-time evolution using\\
forward and backward real-time evolution with a single ancilla:\\
First-quantized eigensolver algorithm for quantum chemistry
}

\author{Taichi Kosugi}
\email{kosugi.taichi@gmail.com}
\author{Yusuke Nishiya}
\author{Hirofumi Nishi}
\author{Yu-ichiro Matsushita}
\affiliation{
Laboratory for Materials and Structures,
Institute of Innovative Research,
Tokyo Institute of Technology,
Yokohama 226-8503,
Japan
}

\affiliation{
Quemix Inc.,
Taiyo Life Nihombashi Building,
2-11-2,
Nihombashi Chuo-ku, 
Tokyo 103-0027,
Japan
}

\date{\today}

\begin{abstract}
Imaginary-time evolution (ITE) on a quantum computer is a promising formalism for obtaining the ground state of a quantum system. 
The probabilistic ITE (PITE) exploits measurements to implement nonunitary operations
and it can avoid the restriction of dynamics to a low-dimensional subspace imposed by variational parameters unlike other types of ITE.
In this study,
we propose a new PITE approach that uses only one ancillary qubit.
Unlike the existing PITE approaches,
the new one under a practical approximation constructs the circuit
from forward and backward real-time evolution (RTE) gates as black boxes for the original Hamiltonian.
Thus, all efficient unitary algorithms for RTE 
can be transferred to the ITE without any modifications.
Our approach can be used to obtain the Gibbs state at a finite temperature and partition function.
We validate the approach via several illustrative systems 
where the trial states are found to converge rapidly to the ground states.
In addition, we discuss its applicability to quantum chemistry by focusing on the scaling of computational cost; this leads to the development of a novel framework referred to as a first-quantized eigensolver.
The nonvariational generic approach will expand the scope of practical quantum computation for versatile objectives. 
\end{abstract}

\maketitle 

\section{Introduction}
\label{sec:introduction}

When solving a practical computational
problem for which quantum parallelism can outperform classical algorithms,
e.g., quantum chemistry\cite{bib:5298} and
combination optimization\cite{bib:4949, bib:5103},
the problem is often rewritten according to the quantum dynamics governed by an appropriately defined many-qubit Hamiltonian.
The problem is, at least computationally, reduced to finding the ground state of the target system.
Imaginary-time evolution (ITE) on classical computers is a technique that is widely used to obtain the ground state of a quantum system.
It is based on the imaginary-time Schr\"odinger equation\cite{bib:5611},
derived by replacing real time $t$ with a purely imaginary value $-i \tau$,
wherein the real parameter $\tau$ is called the imaginary time.
A wave function governed by this equation exhibits dynamics wherein excited states decay rapidly as the imaginary time proceeds.

Recently, there has been a growing interest in ITE for quantum computation.\cite{bib:4797, bib:4802, bib:4807, bib:4959, bib:4980, bib:5613, bib:5612, bib:5260, bib:5600, bib:5568, bib:5409, bib:5566}
However, its implementation is 
not as straightforward as that on a classical computer
because ITE operation is nonunitary,
which makes its direct implementation as a sequence of elementary quantum gates impossible.\cite{Nielsen_and_Chuang} 
There are three main types of indirect but practical implementations of ITE on a quantum computer:
variational ITE (VITE)\cite{bib:4797, bib:4802, bib:4807},
quantum ITE (QITE)\cite{bib:4959, bib:4980, bib:5260, bib:5568, bib:5566},
and probabilistic ITE (PITE).\cite{bib:5600, bib:5409}

The first type, VITE,
solves the imaginary-time Schr\"odinger equation on a classical computer for parameters of unitary ansatz gates.
The equation is derived based on McLachlan's variational principle.\cite{bib:4803}
Further, 
the ansatz circuit must be selected appropriately to attain a converged state that is satisfactorily close to the true ground state because VITE is a type of variational quantum eigensolver (VQE)\cite{bib:4470, bib:4517}.

The second type, QITE,
solves the linear equation on a classical computer to determine the coefficients of Pauli tensors for unitary evolution.
The equation is derived such that 
a unitarily evolved state approximates the exactly (nonunitarily) evolved state in the sense of the squared norm; this derivation is mathematically equivalent to McLachlan's variational principle for VITE.
The matrix dimension for the linear equation tends to grow rapidly with an increase in the problem size,
i.e., the number $n$ of qubits involved in a given system
and that of Pauli tensors for the approximate unitary evolution.
More precisely,
when one decides to expand the approximate unitary operator in terms of Pauli tensors of size $D,$ called the domain size, there exist $4^D$ tensors (including the identity) for each combination of $D$ qubits picked from $n.$\cite{bib:4959}
Naive adoption of such ${}_n \mathrm{C}_D$ combinations leads to a linear equation of dimension $d = 4^D {}_n \mathrm{C}_D.$
Although solving this equation by using conventional methods on a classical computer requires polynomial computational time with respect to $d,$ the time grows rapidly as $D$ and/or $n$ increase.
Its scaling can be regularized if certain approximations that consider the locality of partial Hamiltonians are adopted.\cite{bib:5429}
A QITE circuit can be compressed in several ways.\cite{bib:5429, bib:5567}

The third type, PITE,
exploits measurements to implement nonunitary operations.
An input state undergoes the unitary operations,
and the measurement is performed;
then, the state collapses to the desired state with a certain probability.
Liu et al.\cite{bib:5409} analyzed their PITE circuit for 
the Grover's search algorithm\cite{bib:5462, bib:5463},
in which they introduced an additional qubit per pair of qubits for a nonunitary operation on the pair.
The measurement-based generation of desired states is used for other purposes such as linear equations\cite{bib:4716}, Green's functions\cite{bib:5005, bib:5569},
linear-response functions\cite{bib:5163},
and ionization energies.\cite{bib:5570}
The three types of ITE summarized above are not necessarily exclusive to each other.
In fact, a combination of QITE and PITE was proposed recently.\cite{bib:5557}

Although the implementation of PITE by using only one ancillary qubit is mathematically ensured to be possible\cite{bib:5175, bib:5163, bib:5600},
it requires classical computational resources for the singular-value decomposition of the total many-qubit Hamiltonian or those for other methods having the same cost,
which is typically estimated to be $\mathcal{O} ((2^n)^3)$ for an $n$-qubit system.
In this study,
we propose a generic nonvariational approach of PITE that uses only one ancillary qubit
and that does not demand explicitly large classical cost for linear algebra. 
Within the first-order of the time step,
our approach enables one to construct the circuit from forward and backward real-time evolution (RTE)
generated by the Hamiltonian without knowing its eigenvalues and eigenvectors.
Thus, every unitary quantum algorithm for real-time dynamics
can be transferred to the imaginary-time dynamics without modifications.
Further, our approach can be used to obtain the Gibbs state at a finite temperature and partition function.
We apply the approach to several systems to confirm its validity.
In addition, we discuss the overall PITE procedure of quantum chemistry calculations for obtaining the ground states of molecules by focusing on the scaling of computational cost, which leads to a novel framework referred to as a first-quantized eigensolver (FQE).

\section{Methods}
\label{sec:methods}

\subsection{PITE for generic cases}
\label{sec:methods:PITE_for_generic}

\subsubsection{Exact circuit}

Let $\mathcal{H}$ be the Hamiltonian for an $n$-qubit system.
The ITE operator for an imaginary-time step $\Delta \tau$ is $e^{-\mathcal{H} \Delta \tau}$.
For an input state $| \psi \rangle$,
we want to get the evolved state $e^{-\mathcal{H} \Delta \tau} | \psi \rangle$ up to a normalization constant.
From the evolution operator and an adjustable real parameter $m_0$,
we define a nonunitary Hermitian operator
$
\mathcal{M}
\equiv
    m_0
    e^{-\mathcal{H} \Delta \tau}
    ,
$
for which we impose conditions $0 < m_0 < 1$ and $m_0 \ne 1/\sqrt{2}$.
These conditions allow us perform Taylor expansion safely to obtain
Eq.~(\ref{imag_evol_as_part_of_real_evol:C_ITE_is_approximate_C_ITE_1})
later.
To establish the exact PITE,
we define the following Hermitian operator formally for the $n$-qubit system:
\begin{gather}
    \Theta
    \equiv
        \arccos
        \frac{\mathcal{M} + \sqrt{1 - \mathcal{M}^2}}{\sqrt{2}}
        .
    \label{imag_evol_as_part_of_real_evol:def_Theta}
\end{gather}
This definition with
$\kappa \equiv \mathrm{sgn} ( m_0 - 1/\sqrt{2} )$
leads to
$
\cos \Theta
=
(\mathcal{M} + \sqrt{1 - \mathcal{M}^2})/\sqrt{2}
$
and
$
\sin \Theta
=
\kappa
(\mathcal{M} - \sqrt{1 - \mathcal{M}^2})/\sqrt{2}
.
$
The operators $e^{\pm i \kappa \Theta}$ are unitary
and hence they can be implemented as quantum gates for the $n$ qubits.

\begin{figure}
\begin{center}
\includegraphics[width=8.5cm]{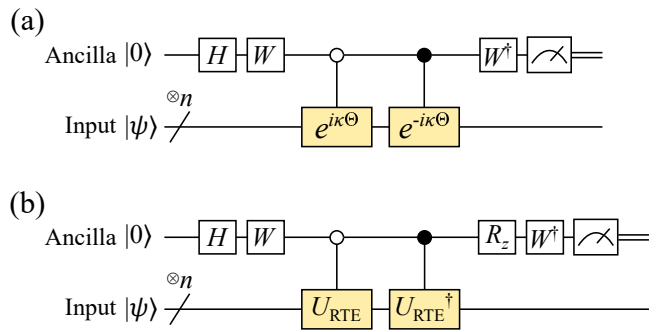}
\end{center}
\caption{
(a)
Circuit $\mathcal{C}_{\mathrm{PITE}}$
for probabilistic preparation of the exact imaginary-time-evolved state of an input $n$-qubit state $| \psi \rangle$.
$H$ is the Hadamard gate.
The yellow boxes represent the gates involving the Hamiltonian.
If the measurement outcome is $| 0 \rangle$,
the input state has collapsed to the desired state.
(b)
Approximate circuit $\mathcal{C}_{\mathrm{PITE}}^{(1)}$,
equivalent to $\mathcal{C}_{\mathrm{PITE}}$
within the first order of $\Delta \tau$.
$U_{\mathrm{RTE}} \equiv U_{\mathrm{RTE}} (s_1 \Delta \tau)$
and
$R_z \equiv R_z (-2 \theta_0)$
are used in this figure.
}
\label{circuit:imag_evol_as_part_of_real_evol:evolution_exact_and_st1}
\end{figure}

We introduce an ancillary qubit to construct the circuit $\mathcal{C}_{\mathrm{PITE}}$
for the $(n + 1)$-qubit system, as shown in
Fig.~\ref{circuit:imag_evol_as_part_of_real_evol:evolution_exact_and_st1}(a).
The circuit uses the single-qubit gate
\begin{gather}
    W
    \equiv
        \frac{1}{\sqrt{2}}
        \begin{pmatrix}
            1 & -i \\
            1 & i
        \end{pmatrix}
        .
    \label{imag_evol_as_part_of_real_evol:def_W}
\end{gather}
The composite system undergoes the unitary operations as
\begin{gather}
    | \psi \rangle
    \otimes
    | 0 \rangle
    \longmapsto{}
            \mathcal{M}
            | \psi \rangle
            \otimes
            | 0 \rangle
            +
            \sqrt{1 - \mathcal{M}^2}
            | \psi \rangle
            \otimes
            | 1 \rangle
            ,
    \label{imag_evol_as_part_of_real_evol:whole_state_before_measurement}
\end{gather}
which is the state immediately before the measurement on the ancilla.
(See also Appendix \ref{sec:state_change_during_exact_circuit}.)
We denote the $n$-qubit state 
$\mathcal{M} | \psi \rangle \propto e^{-\mathcal{H} \Delta \tau} | \psi \rangle$,
coupled to the ancillary state $| 0 \rangle$ in the entangled state on the RHS of 
Eq.~(\ref{imag_evol_as_part_of_real_evol:whole_state_before_measurement}),
by the success state in what follows.
This state is nothing but the desired imaginary-time-evolved state.
We denote, on the other hand, the state $\sqrt{1 - \mathcal{M}^2} | \psi \rangle$,
coupled to the ancillary state $| 1 \rangle$,
by the failure state in what follows.
The Born's rule tells us that we will get the success state
if the measurement outcome is $| 0 \rangle$,
with the probability
$
\mathbb{P}_0
=
\langle \psi | \mathcal{M}^2 | \psi \rangle
.
$
The PITE is thus written symbolically as
$
| \psi \rangle
\xmapsto{\mathrm{If \ success}}
\mathcal{M} | \psi \rangle/\sqrt{\mathbb{P}_0}
,
$
where the normalization constant is correctly taken into account.

\subsubsection{Approximate circuit}

It is known\cite{bib:5175} that an arbitrary diagonalizable complex matrix can be written as a linear combination of at most two unitary matrices provided that the resources of classical computation for its singular-value decomposition is available.
The probabilistic implementation of a linear combination of untaries\cite{bib:5163} together with this fact ensures mathematically the possibility of probabilistic imaginary-time dynamics with a single ancillary qubit (see also Ref.~\cite{bib:5600}).
Since the exact PITE described above involves no approximation and is applicable to an arbitrary $\Delta \tau$,
it is mathematically desirable.
The adoption of it, however, is hardly of practical use due to the operator $\Theta$.
Specifically, it is formidable to implement the exponentiated $\Theta$ as a sequence of elementary quantum gates for the multiple qubits without knowing its eigenvalues and eigenvectors.
Therefore we have to circumvent such a difficulty.

As usual, we assume that the imaginary-time step is small.
We can Taylor expand $\Theta$,
defined by Eq.~(\ref{imag_evol_as_part_of_real_evol:def_Theta}),
in $\Delta \tau$ for such a case as
$
\kappa
\Theta
=
    \theta_0
    -
    \mathcal{H}
    s_1 \Delta \tau
    +
    \mathcal{O} (\Delta \tau^2)
,
$
where
$
\theta_0
\equiv
\kappa
\arccos [(m_0 + \sqrt{1 - m_0^2})/\sqrt{2}]
$
and
$s_1 \equiv m_0/\sqrt{1 - m_0^2}.$
We can rewrite the anti-controlled $e^{i \kappa \Theta}$
and the subsequent controlled $e^{-i \kappa \Theta}$
in the exact circuit in Fig.~\ref{circuit:imag_evol_as_part_of_real_evol:evolution_exact_and_st1}(a) as
\begin{gather}
    e^{i \kappa \Theta}
    \otimes
    | 0 \rangle \langle 0 |
    +
    e^{-i \kappa \Theta}
    \otimes
    | 1 \rangle \langle 1 |
    \nonumber \\
    =
        \Big(
            I_{2^n}
            \otimes
            R_z (-2 \theta_0)
        \Big)
        \Big(
            U_{\mathrm{RTE}} (s_1 \Delta \tau)
            \otimes
            | 0 \rangle \langle 0 |
    \nonumber \\
            +
            U_{\mathrm{RTE}} (s_1 \Delta \tau)^\dagger
            \otimes
            | 1 \rangle \langle 1 |
        \Big)
        +
        \mathcal{O} (\Delta \tau^2)
        ,
    \label{imag_evol_as_part_of_real_evol:C_ITE_is_approximate_C_ITE_1}
\end{gather}
where $I_{2^n}$ is the identity for an $n$-qubit state and
$R_z (\theta)$ is the single-qubit $z$ rotation.
$
U_{\mathrm{RTE}} (\Delta t)
\equiv
e^{-i \mathcal{H} \Delta t}
$
is the usual RTE operator with respect to the original Hamiltonian
for a real-time step $\Delta t$.
By using Eq.~(\ref{imag_evol_as_part_of_real_evol:C_ITE_is_approximate_C_ITE_1}),
we can construct the circuit $\mathcal{C}_{\mathrm{PITE}}^{(1)}$ as shown in
Fig.~\ref{circuit:imag_evol_as_part_of_real_evol:evolution_exact_and_st1}(b),
which implements approximately the exact one, $\mathcal{C}_{\mathrm{PITE}}$,
within the first order of $\Delta \tau$.
In fact, it is easily confirmed that 
the state of the $(n + 1)$-qubit system immediately before the measurement is 
\begin{gather}
        m_0
        (1 - \mathcal{H} \Delta \tau)
        | \psi \rangle
        \otimes
        | 0 \rangle
    \nonumber \\
        +
        \left( \sqrt{1 - m_0^2} + \frac{m_0^2}{\sqrt{1 - m_0^2}} \mathcal{H} \Delta \tau \right)
        | \psi \rangle
        \otimes
        | 1 \rangle
        +
        \mathcal{O} (\Delta \tau^2)
        ,
    \label{imag_evol_as_part_of_real_evol:whole_state_before_measurement_tau_1}
\end{gather}
consistent with the RHS of
Eq.~(\ref{imag_evol_as_part_of_real_evol:whole_state_before_measurement})
within the first order of $\Delta \tau$.
The forward
$U_{\mathrm{RTE}} (s_1 \Delta \tau)$
and backward
$U_{\mathrm{RTE}} (s_1 \Delta \tau)^\dagger$
RTE operators for the rescaled time step
in the circuit are responsible for the imaginary-time dynamics.
The approximate circuit found here is the main result of the present work.

The circuit $\mathcal{C}_{\mathrm{PITE}}^{(1)}$ is suggestive of sophisticated development of quantum algorithms for ground states of quantum systems.
The direct operations on the input register come solely as the RTE generated by the Hamiltonian,
which means that we have pushed the intricacy about the nonunitary operation
$e^{-\mathcal{H} \Delta \tau}$ into the two black boxes, or equivalently an oracle,
in which the tractable unitary operations of the form $e^{\pm i \mathcal{H} \Delta t}$ are implemented.
Every efficient unitary algorithm for the real-time dynamics can thus be transferred to the imaginary-time dynamics exactly as it is.
This availability is favorable since the real-time dynamics in quantum computation is much more tractable than the imaginary-time dynamics in general.

Since the accuracy of evolution is worse for a larger $\Delta \tau$,
we need to apply the approximate circuit for a small $\Delta \tau$ as a Trotter step and perform the measurements repeatedly.
We have to get the success states for all the measurements to reach the final state where all the excited states have diminished sufficiently.
It is easily understood for an $n_{\mathrm{steps}}$-step evolution that
the probability for obtaining the success state having survived
all the $n_{\mathrm{steps}}$ measurements is
$\langle \psi | \mathcal{M}^{2 n_{\mathrm{steps}}} | \psi \rangle$,
exponentially decaying as the measurement number increases.

The construction of our single-ancilla PITE circuit is possible primarily due to the Hermiticity of the generator for evolution.
For the dynamics generated by a non-Hermitian operator,
the probabilistic evolution is also possible if two ancillary qubits are available.
For details, see Appendix \ref{sec:prob_evol_for_non_Hermitian}.

\subsubsection{Comparison with other ITE approaches}

The advantage and disadvantage of PITE compared with VITE and QITE have to be mentioned.
First of all, PITE does not involve arbitrariness for the implementation of state evolution of qubits on a circuit.
That is,
PITE does not need an ansatz circuit unlike VITE and
it does not need selection of Pauli tensors according to the locality of partial Hamiltonians unlike QITE.
The expected energy in a PITE calculation is thus minimized in the full Hilbert space of qubits which encode the state of target system.
On the other hand,
the exponential decrease in success probability during the PITE steps is a deplorable drawback inherent to the PITE approach
as long as we are not interested in the normalization constant.
A promising alternative for alleviating this difficulty is the quantum amplitude amplification (QAA) technique,\cite{bib:4884, bib:4878} proposed by Brassard and his coworkers,
known as a generalization of the Grover's search algorithm.\cite{bib:5462, bib:5463}
In the original QAA,\cite{bib:4878}
the rotation angle within the subspace is determined by the quantum amplitude estimation (QAE),
which is a kind of quantum phase estimation (QPE).\cite{Nielsen_and_Chuang}
It is possible, however, to perform QAA without QAE by employing the maximum likelihood estimation, as demonstrated by Suzuki et al.\cite{bib:5145}
The examinations of PITE in combination with QAA will be interesting and important for practical applications of the PITE approach where the input state has only a small overlap with the ground state.

\subsection{FQE framework for quantum chemistry}
\label{sec:methods:PITE_for_real_space}

One of the most practical applications for finding the ground states of quantum systems is quantum chemistry calculations for molecules,
where many electrons interact with each other in three-dimensional space.
As a promising application of our PITE approach,
we describe here the framework for obtaining the ground state wave function of a molecular system.
This framework, which we call FQE, is based on the scheme proposed by Kassal et al.\cite{bib:5328} for efficient implementation of real-time dynamics for quantum chemistry.

\subsubsection{Single particle in one-dimensional space}

We begin with the consideration of a single particle in one-dimensional space as the first step for establishing the FQE framework.
We assume that the particle with a mass $m$ moves within a range $[0, L]$ on the $x$ axis,
whose first-quantized Hamiltonian is $\mathcal{H} = \hat{T} + V (\hat{x}).$
$\hat{T} \equiv \hat{p}^2/(2 m)$ is the kinetic-energy operator for the momentum operator $\hat{p}.$
$V$ is the external potential for the position operator $\hat{x}$. 
We assume that $n$ qubits are provided us for representing the wave function in real space.

We want to implement the RTE operator $e^{-i \mathcal{H} \Delta t}$ for a real-time step $\Delta t$
that acts on the qubits whose state encodes the wave function of the particle.
We discretize the finite interval to define $N \equiv 2^n$ equidistant points,
that is,
$x^{(k)} \equiv k \Delta x \ (k = 0, \dots, N - 1)$
for the step $\Delta x \equiv L/N.$
We can thus assign each computational basis $| k \rangle_n$ to each of the position eigenstates:
$\hat{x} | k \rangle_{n} = x^{(k)} | k \rangle_{n}.$
The coefficient of each computational basis in an $n$-qubit state thus represents the wave function $\psi (x)$ of the particle as
$
| \psi \rangle
=
\sqrt{\Delta x}
\sum_{k = 0}^{N - 1}
    \psi(x^{(k)})
    | k \rangle_n
,
$
where $\sqrt{\Delta x}$ is for normalization.

As the canonical counterpart of the discretized positions,
we define the $N$ discrete momenta
$
p^{(\widetilde{s})}
\equiv
\widetilde{s}
\Delta p
\
(\widetilde{s} = -N/2, -N/2 + 1 \dots, N/2 - 1)
$
for the step $\Delta p \equiv 2 \pi /L$ in reciprocal space.
Recalling that the quantum field theory\cite{fetter2003quantum, stefanucci2013nonequilibrium} relates
the creation and annihilation operators of momentum eigenstates
with those of position eigenstates via Fourier transform,
we should define in this case the momentum eigenstate as
\begin{gather}
    | p^{(\widetilde{s})} \rangle
    \equiv
        \frac{1}{\sqrt{N}}
        \sum_{k = 0}^{N - 1}
        \exp \left( i p^{(\widetilde{s})} x^{(k)} \right)
        | k \rangle_n
        .
    \label{QITE_as_a_part_of_RITE:def_mom_eigenstate}
\end{gather}
We have to notice, however, that
these eigenstates of position and momentum do not transform between each other via the ordinary quantum Fourier transform (QFT).\cite{Nielsen_and_Chuang}
That is, the minimum of position is $0$, while that of momentum is $-N \Delta p/2$.
This discrepancy is taken into account appropriately by employing the centered QFT (CQFT)\cite{bib:5391, bib:5384},
which is defined as the application of QFT after $\sigma_x$ on the qubit corresponding to the highest bit for the index of an $n$-qubit computational basis:
$
\mathrm{CQFT}
=
\mathrm{QFT}
\left( \sigma_x \otimes I_{2^{n - 1}} \right)
.
$
One can easily confirm that CQFT transforms the position eigenstate to the momentum eigenstate correctly:
$
\mathrm{CQFT}
| k \rangle_{n}
=
| p^{(\widetilde{k})} \rangle
$
,
where
$\widetilde{k} \equiv k - N/2.$
The meaning of the tilde symbol for an integer is the same as this in what follows.

$| p^{(\widetilde{s})} \rangle$ is also the eigenstate of $\hat{T}$ belonging to the discrete kinetic energy
$E_s \equiv \widetilde{s}^2 (\Delta p)^2/(2 m).$
If we have a kinetic-phase gate $U_{\mathrm{kin}}$
which acts on the computational basis diagonally as
$
U_{\mathrm{kin}} (\Delta t) | j \rangle_n
\equiv
\exp ( -i E_j \Delta t ) | j \rangle_n
,
$
it becomes under CQFT as
\begin{align}
    \mathrm{CQFT}
    \cdot
    U_{\mathrm{kin}} (\Delta t)
    \cdot
    \mathrm{CQFT}^\dagger
    &=
        \sum_{s = 0}^{N - 1}
            e^{-i E_s \Delta t}
            | p^{(\widetilde{s})} \rangle
            \langle p^{(\widetilde{s})} |
    \nonumber \\
    &=
        e^{-i \hat{T} \Delta t}
        ,
    \label{imag_evol_as_part_of_real_evol:CQFT_on_kinetic_rotation}
\end{align}
where we used the completeness of the momentum eigenstates to get the last equality.
Equation (\ref{imag_evol_as_part_of_real_evol:CQFT_on_kinetic_rotation})
means that the evolution coming from the kinetic operator can be implemented by using CQFT and the kinetic-phase gate.\cite{bib:5384}
For the details of the action of kinetic-evolution operator on the position eigenstate,
see Appendix~\ref{sec:kinetic_propagator}.

As for the potential-evolution operator $e^{-i V (\hat{x}) \Delta t},$
we need a potential-phase gate $U_{\mathrm{pot}}$
which acts on the position eigenstate diagonally as
$
U_{\mathrm{pot}} (\Delta t) | k \rangle_n
\equiv
\exp ( -i V (x^{(k)}) \Delta t ) | k \rangle_n
.
$
There exist several alternatives for implementation of the kinetic- and potential-phase gates.
While Kassal et al. in the original paper\cite{bib:5328}
adopted the phase kick-back technique\cite{bib:5393},
Ollitrault et al.\cite{bib:5384} adopted the technique proposed by 
Benenti and Strini\cite{bib:5390},
which introduces to a computational basis the phase factor given as a polynomial of position.
The extension of the latter technique for a piecewisely defined polynomial is possible\cite{bib:5384} in combination with efficient comparators\cite{bib:5389},
which decide the interval the index of an input computational basis falls onto.

The kinetic energy and potential operators do not commute with each other in most systems.
Therefore we have to consider application of them to the qubits sequentially by employing the first-order Suzuki--Trotter as
$
e^{-i \mathcal{H} \Delta t}
=
e^{-i \hat{T} \Delta t} e^{-i V(\hat{x}) \Delta t} +\mathcal{O} (\Delta t^2)
$
for the RTE.
Given this approximation, we adopt the circuit $\mathcal{C}_{\mathrm{PITE}}^{(1)}$ in
Fig.~\ref{circuit:imag_evol_as_part_of_real_evol:evolution_exact_and_st1}(b)
for PITE.
Since the RTE operator in this case is
$
U_{\mathrm{RTE}} (s_1 \Delta \tau)
=
    \mathrm{CQFT}
    \cdot
    U_{\mathrm{kin}} (s_1 \Delta \tau)
    \cdot
    \mathrm{CQFT}^\dagger
    \cdot
    U_{\mathrm{pot}} (s_1 \Delta \tau)
,
$
the neighboring CQFT and CQFT$^\dagger$ in the forward and backward evolution fortunately cancel each other,
leading to the circuit $\mathcal{C}_{\mathrm{PITE}}^{(\mathrm{ST1})}$ shown in
Fig.~\ref{fig:pite_circuit_for_1dim_and_electrons}(a).
The number of operations for each imaginary-time step is
$
c_{\mathrm{PITE}}
\approx
2 
( c_{\mathrm{CQFT}} + \alpha c_{\mathrm{kin}} + \alpha' c_{\mathrm{pot}} )
,
$
where 
$c_{\mathrm{CQFT}} = \mathcal{O}(n^2)$ is that of CQFT.\cite{Nielsen_and_Chuang}
$c_{\mathrm{kin}}$ and $c_{\mathrm{pot}}$ are those of
the kinetic- and potential-phase gates, respectively.
$\alpha$ and $\alpha'$ are constants independent of $n$,
coming from the introduction of control bits.
If we adopt the implementation of phase gates in Refs.\cite{bib:5390, bib:5384},
the operation number of the kinetic-phase gate is
$c_{\mathrm{kin}} = \mathcal{O} (n^2)$.
Similarly, if the potential is given as a single or a piecewisely defined $p$th-order polynomial, the operation number of the potential-phase gate is
$c_{\mathrm{pot}} = \mathcal{O} (n^p)$.
The operation number for the PITE step is thus
$c_{\mathrm{PITE}} = \mathcal{O} (\max (n^2, n^p)).$

The analyses of operation numbers in the circuits provided in this paper should be mentioned here.
If one aims to discuss strictly the operation number of a given circuit,
an elementary gate set has to be defined so that each member is the unit for counting the operations.
A typical elementary gate set consists of a small number of single-qubit gates and a CNOT gate.\cite{Nielsen_and_Chuang}
In the present study, however, we do not need to introduce any specific elementary gate set for discussing the scaling of operation numbers with respect to the numbers of grid points and particles.
It is because every many-qubit operation appearing in this study is a controlled single-qubit operation for which the number of control qubits is independent of the numbers of grid points and particles.
Specifically,
each QFT gate is implemented by using singly controlled rotations, CNOTs, and SWAPs\cite{Nielsen_and_Chuang}.
The depth of each of these building blocks
is independent of the grid points and the particle number.
This is similarly the case for the phase gates comprising the entire PITE circuit since we implement them with the techniques in Refs.\cite{bib:5390, bib:5384}

\subsubsection{Interacting electrons in three-dimensional space}

Let us move on to establishing the FQE framework.
We discuss here the encoding of a many-electron wave function
and 
the construction of PITE circuit for a quantum chemistry problem,
based on which we estimate the scaling of computational cost as a function of problem size.
The discussion below defines and characterizes the FQE framework.
We ignore the spin degree of freedom for simplicity in what follows.

The extension of the scheme for the case of a single particle in one-dimensional space explained above to the case of interacting electrons in three-dimensional space is straightforward.
Let us consider $n_{\mathrm{el}}$ electrons confined to a cube as a simulation cell,
each of whose edges has a length $L$.
We assume that $n_{\mathrm{d}}$ qubits are available for the degree of freedom in each direction for each electron.
We discretize the simulation cell to define a regular mesh which has
$N_{\mathrm{d}} \equiv 2^{n_{\mathrm{d}}}$ points in each direction.
The wave function for an electron can thus be encoded in $3 n_{\mathrm{d}}$ qubits,
whose coefficients of computational basis represent the wave function at
$N_{\mathrm{grid}} \equiv N_{\mathrm{d}}^3$ equidistant grid points.
The $n_{\mathrm{el}}$-electron wave function
$\Psi (\boldsymbol{r}_1, \dots, \boldsymbol{r}_{n_{\mathrm{el}}})$
is thus encoded in $3 n_{\mathrm{d}} n_{\mathrm{el}}$ qubits
as a linear combination of 
$
| k_{1 x}, k_{1 y}, k_{1 z} \rangle 
\otimes \cdots \otimes
| k_{n_{\mathrm{el}} x}, k_{n_{\mathrm{el}} y}, k_{n_{\mathrm{el}} z} \rangle 
,
$
where
$| k_{\mu x}, k_{\mu y}, k_{\mu z} \rangle$
is the position eigenstate of $\mu$th electron specified by the three integers.
The required memory for storing the wave function on a classical computer is
$\mathcal{O} ( (N_{\mathrm{grid}})^{n_{\mathrm{el}}})$,
scaling exponentially with respect to $n_{\mathrm{el}}$ and $n_{\mathrm{d}}.$
On the other hand, the required number of qubits for the storage is
$\mathcal{O} (n_{\mathrm{el}} \log N_{\mathrm{grid}})$,
exhibiting polynomial scaling,
that is the fundamental advantage of quantum simulation discussed by Feynman.\cite{bib:4828}
The extension of our formalism described below to a case in which the simulation cell is spanned by arbitrary three independent vectors will be straightforward.
In addition, the numbers of qubits for encoding the wave functions can differ from each other for the three directions.

The first-quantized Hamiltonian for electrons involves typically interactions of the form
$
\hat{V}_{\mathrm{int}}
=
\sum_{\mu > \mu'}
v_{\mathrm{int}} (\hat{\boldsymbol{r}}_{\mu}, \hat{\boldsymbol{r}}_{\mu'})
,
$
where $v_{\mathrm{int}}$ is a pairwise interaction such as the Coulomb interaction.
The implementation of RTE thus requires the interaction-phase gate
$U_{\mathrm{int}}$ that acts diagonally on the position eigenstate for a pair of electrons as
$
U_{\mathrm{int}} (\Delta t)
| \boldsymbol{r} \rangle
\otimes
| \boldsymbol{r}' \rangle
\equiv
\exp
(- i v_{\mathrm{int}} (\boldsymbol{r}, \boldsymbol{r}') \Delta t ) 
| \boldsymbol{r} \rangle
\otimes
| \boldsymbol{r}' \rangle
.
$
If $v_{\mathrm{int}}$ is given as a simple or a piecewisely defined $p'$th-order polynomial,
the evolution operator
$\exp (-i \hat{V}_{\mathrm{int}} \Delta t)$
can be implemented with the operation number 
$c_{\mathrm{int}} = \mathcal{O} ( n_{\mathrm{el}}^2 n_{\mathrm{d}}^{p'})$
by using the techniques in Refs.\cite{bib:5390, bib:5384}

\begin{figure*}
\begin{center}
\includegraphics[width=17cm]{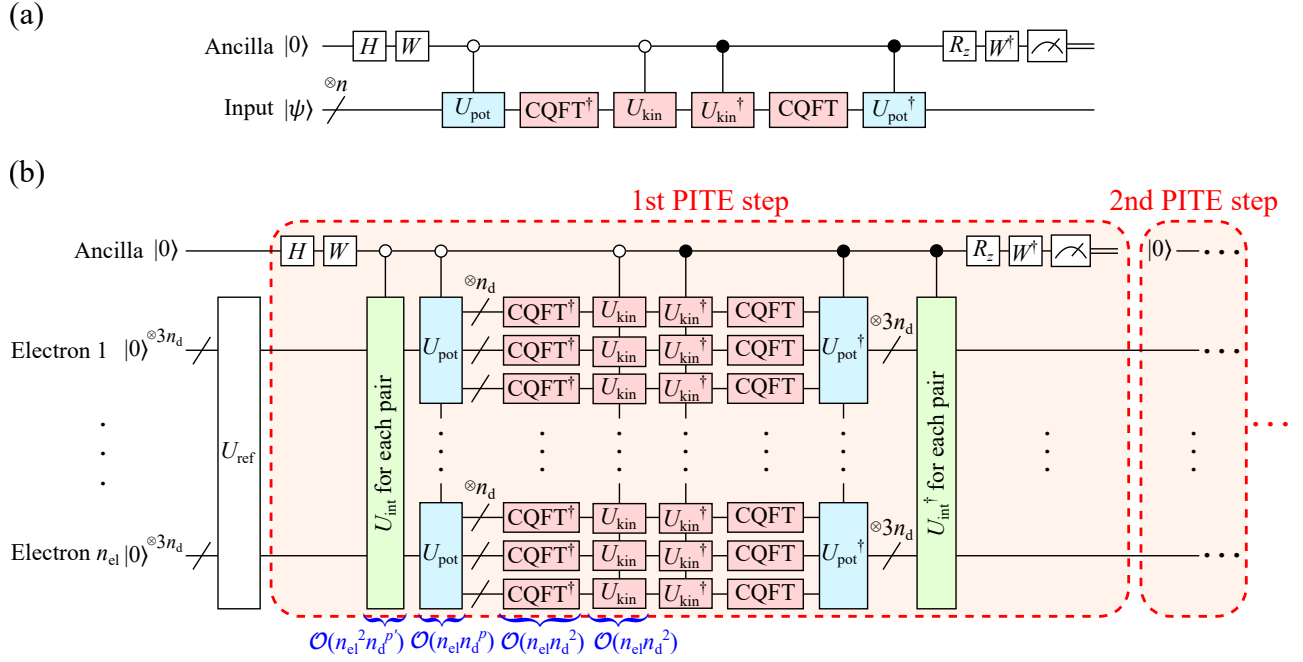}
\end{center}
\caption{
(a)
Circuit $\mathcal{C}_{\mathrm{PITE}}^{(\mathrm{ST1})}$ as a special case of
$\mathcal{C}_{\mathrm{PITE}}^{(1)}$ in
Fig.~\ref{circuit:imag_evol_as_part_of_real_evol:evolution_exact_and_st1}(b).
This circuit implements PITE for a particle in one-dimensional real space within the first-order Suzuki--Trotter.
$U_{\lambda} \equiv U_{\lambda}(s_1 \Delta \tau) \ (\lambda = \mathrm{kin, pot}),$
defined in the main text,
is used in this figure.
(b)
FQE procedure for interacting electrons in three-dimensional space.
$3 n_{\mathrm{d}}$ qubits are used for each of the $n_{\mathrm{el}}$ electrons to encode the many-electron wave function.
$U_{\mathrm{ref}}$ prepares the reference state in which the Fermi statistics of electrons is taken into account.
The red dashed rectangles represent the PITE steps.
The operation numbers for the individual components in the PITE step are shown in blue.
}
\label{fig:pite_circuit_for_1dim_and_electrons}
\end{figure*}

The whole FQE procedure for the interacting electrons is shown in 
Fig.~\ref{fig:pite_circuit_for_1dim_and_electrons}(b).
$U_{\mathrm{ref}}$ prepares the reference state as the input to the first PITE step.
We have to keep in mind that the reference state has to be generated carefully
so that it is antisymmetric under arbitrary exchange of the electrons.\cite{bib:4825}
Berry et al.\cite{bib:5389} proposed an efficient technique for
antisymmetrization of a many-electron wave function.
Such a process is needed in general when we work with the first-quantized formalism,
in contrast to the second-quantized one.
The kinetic-evolution operators for each electron can be implemented for the three directions independently, as shown in the figure.
The operation numbers for the individual components in the PITE step are also shown.
With possibly $p' \geq 2$ in practice,
the largest number $c_{\mathrm{int}}$ of operations
comes from $U_{\mathrm{int}}$ [see Fig.~\ref{fig:pite_circuit_for_1dim_and_electrons}(b)],
which dominates the overall scaling of the computational cost $c_{\mathrm{PITE}}$
of the whole PITE step for increase in the electron number.

The accuracy of quantum dynamics depends on
the resolution of the wave function, or equivalently the step $\Delta x$
between neighboring grid points in each direction.
Since the number of electrons is roughly proportional to the volume of simulation cell in a typical calculation,
we have, for a fixed grid step,
$
n_{\mathrm{d}} \propto \log (L/\Delta x)
\propto
\log (n_{\mathrm{el}}^{1/3}/\Delta x).
$
The required number of qubits for the many-electron wave function is thus estimated to be
$\mathcal{O} (n_{\mathrm{el}} \log (n_{\mathrm{el}}^{1/3}/\Delta x)).$
The operation number in the whole PITE step in this case is
$
c_{\mathrm{PITE}}
=
\mathcal{O} (n_{\mathrm{el}}^2 (\log (n_{\mathrm{el}}^{1/3}/\Delta x))^{p'} ).
$

If we adopted the second-quantized formalism like VQE instead of the first-quantized one,
we had to work with the creation and annihilation operators of electrons for the molecular orbitals
and hence the circuit in
Fig.~\ref{fig:pite_circuit_for_1dim_and_electrons}(b)
could not be used.
We have to instead construct the circuit in
Fig.~\ref{circuit:imag_evol_as_part_of_real_evol:evolution_exact_and_st1}(b) naively for the second-quantized Hamiltonian.
For $n_{\mathrm{orbs}}$ localized orbitals at each atom in a target molecule,
$\mathcal{O}(n_{\mathrm{orbs}} n_{\mathrm{el}})$ qubits are required for representing the many-electron state.
Due to the numerous two-electron integrals in the Hamiltonian\cite{Helgaker},
$c_{\mathrm{PITE}}$ in this case is at least
$\mathcal{O} ((n_{\mathrm{orbs}} n_{\mathrm{el}})^4).$
The first-quantized formalism is thus more favorable than the second-quantized one in order for a real quantum computer to conform to the coherence time.
Furthermore, the classical computation of the two-electron integrals between molecular-orbitals is needed with cost $\mathcal{O} ((n_{\mathrm{orbs}} n_{\mathrm{el}})^5)$
before the quantum computation starts.
The FQE framework is free from such an overhead and it is hence a promising candidate for quantum chemistry calculation of the ground states of large molecules.

\subsection{PITE for finite-temperature states}
\label{sec:methods:PITE_for_finite_temperatures}

The thermodynamic property of a system subject to a heat bath is described by the partition function
$Z (\beta)$ for an inverse temperature $\beta$.
The thermal equilibrium is attained so that the Helmholtz free energy
$F$ is minimized.
It is related with the partition function as
$e^{-\beta F (\beta)} =  Z (\beta).$
Various approaches for treating finite-temperature states on a quantum computer have been proposed.\cite{bib:5582, bib:4807, bib:5264, bib:4959, bib:5558, bib:5559, bib:5560, bib:5260, bib:5561, bib:5562, bib:5563}
We demonstrate here that our PITE approach can also generate the Gibbs state and calculate the partition function.

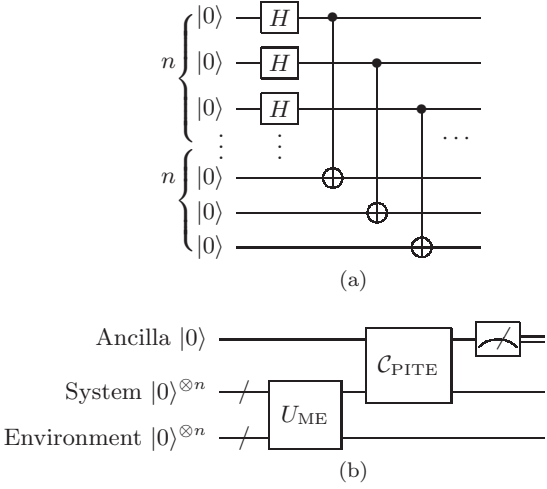
\begin{figure}
\ifnumcomp{\value{is_pdf_used_for_fig}}{=}{1}{
\begin{center}
\end{center}
}{
\centering
\subfloat[]{
\Qcircuit @C=1em @R=.6em {
&&\lstick{| 0 \rangle} & \gate{H} & \ctrl{5} & \qw      & \qw      & \qw & \qw & \\
\lstick{n} &&\lstick{| 0 \rangle} & \gate{H} & \qw      & \ctrl{5} & \qw      & \qw & \qw &  \\
&&\lstick{| 0 \rangle} & \gate{H} & \qw      & \qw      & \ctrl{5} & \qw & \qw & \\
&&\lstick{\vdots}      & \vdots   &          &          &          & \cdots & & &  \\
&&      &     &          &          &          &  & & &  \\
\lstick{n} &&\lstick{| 0 \rangle} & \qw      & \targ    & \qw      & \qw      & \qw & \qw & \\
&&\lstick{| 0 \rangle} & \qw      & \qw      & \targ    & \qw      & \qw & \qw & \\
&&\lstick{| 0 \rangle} & \qw      & \qw      & \qw      & \targ    & \qw & \qw 
\gategroup{1}{1}{4}{1}{.2em}{\{}
\gategroup{5}{1}{8}{1}{.2em}{\{}
} 
}

\subfloat[]{
\Qcircuit @C=1em @R=1em { 
&&\lstick{\mathrm{Ancilla} \ | 0 \rangle}             &     \qw & \qw                            & \multigate{1}{\mathcal{C}_{\mathrm{PITE}}} & \meter &  \cw \\
&&\lstick{\mathrm{System} \ | 0 \rangle^{\otimes n}} & {/} \qw & \multigate{1}{U_{\mathrm{ME}}} & \ghost{\mathcal{C}_{\mathrm{PITE}}}        & \qw    &   \qw \\
&&\lstick{\mathrm{Environment} \ | 0 \rangle^{\otimes n}} & {/} \qw & \ghost{U_{\mathrm{ME}}}        & \qw                                        & \qw    & \qw
} 
}
}
\caption{
(a)
$2 n$-qubit gate $U_{\mathrm{ME}}$
for generating the maximally entangled state from
$| 0 \rangle^{\otimes n} \otimes | 0 \rangle^{\otimes n}.$
(b)
$(2 n + 1)$-qubit circuit $\mathcal{C}_{\mathrm{Gibbs}}$
for probabilistic preparation of the Gibbs state of an $n$-qubit system.
This circuit consists of $\mathcal{C}_{\mathrm{PITE}}$ in
Fig.~\ref{circuit:imag_evol_as_part_of_real_evol:evolution_exact_and_st1}(a)
for the target system and $U_{\mathrm{ME}}.$
If the measurement outcome is $| 0 \rangle$,
the target system has collapsed to the Gibbs state.
The success probability gives the partition function.
}
\label{circuit:max_entangled_state_and_Gibbs} 
\end{figure}

The partition function is defined as
$
Z (\beta)
\equiv
\mathrm{Tr}_{\mathrm{sys}} \exp(-\beta \mathcal{H})
,
$
where the trace is for the $n$-qubit target system having the Hamiltonian $\mathcal{H}$.
We want to prepare the Gibbs state $e^{-\beta \mathcal{H}}/Z(\beta).$
To this end, we use the maximally entangled $2 n$-qubit state
generated by the $U_{\mathrm{ME}}$ gate defined in
Fig.~\ref{circuit:max_entangled_state_and_Gibbs}(a).
This gate entangles the input $n$ pairs of initialized qubits to make them
the Bell pairs.\cite{Nielsen_and_Chuang}
We construct the circuit $\mathcal{C}_{\mathrm{Gibbs}}$
for probabilistic preparation of the Gibbs state,
as shown in Fig.~\ref{circuit:max_entangled_state_and_Gibbs}(b).
This circuit contains $U_{\mathrm{ME}}$ as a part of it to generate
the maximally entangled state between the target system and the environment composed of the same number of qubits.
We set the imaginary-time step for $\mathcal{C}_{\mathrm{PITE}}$ to $\Delta \tau \equiv \beta/2.$
The composite system undergoes the unitary operations as
\begin{gather}
    \overbrace{
        | 0 \rangle^{\otimes n}
    }^{\mathrm{Environment}}
    \otimes
    \overbrace{
        | 0 \rangle^{\otimes n}
    }^{\mathrm{System}}
    \otimes
    \overbrace{
        | 0 \rangle
    }^{\mathrm{Ancilla}}
    \nonumber \\
    \longmapsto{}
        \frac{1}{\sqrt{2^n}}
        \sum_{j = 0}^{2^{n - 1}}
            | j \rangle_n
            \otimes
        \left(
            \mathcal{M}
            | j \rangle_n
            \otimes
            | 0 \rangle
            +
            \sqrt{1 - \mathcal{M}^2 }
            | j \rangle_n
            \otimes
            | 1 \rangle
        \right)
    \nonumber \\
    \equiv
        | \Psi \rangle
        ,
\end{gather}
which is the state immediately before the measurement on the ancilla.
$| j \rangle_n \ (j - 0, \dots, 2^n - 1)$
is the computational basis of an $n$-qubit system.
The reduced density operator\cite{Nielsen_and_Chuang} of the $(n + 1)$-qubit system
consisting of the target system and the ancilla is obtained by
tracing out the environmental qubits as
\begin{gather}
    \rho
    =
        \mathrm{Tr}_{\mathrm{env}}
        | \Psi \rangle \langle \Psi |
    \nonumber \\
    =
        \frac{1}{2^n}
        \Bigg(
            \mathcal{M}^2
            \otimes
            | 0 \rangle \langle 0 |
            +
            \mathcal{M}
            \sqrt{1 - \mathcal{M}^2 }
            \otimes
            | 0 \rangle \langle 1 |
    \nonumber \\
            +
            \sqrt{1 - \mathcal{M}^2 }
            \mathcal{M}
            \otimes
            | 1 \rangle \langle 0 |
            +
            \left(
                1 - \mathcal{M}^2
            \right)
            \otimes
            | 1 \rangle \langle 1 |
        \Bigg)
        .
\end{gather}
The probability for obtaining $| 0 \rangle$ from the measurement is thus
\begin{gather}
    \mathbb{P}_0
    =
        \mathrm{Tr}
        (\mathcal{P}_0 \rho)
    =
        m_0^2
        \frac{Z (\beta)}{2^n}
        ,
    \label{PITE_at_finite_temperatures:prob_0}
\end{gather}
where $\mathcal{P}_0$ is the projection operator onto
the subspace spanned by the ancillary $| 0 \rangle$ state.
The $(n + 1)$-qubit state for this outcome will be
$
\mathcal{P}_0 \rho \mathcal{P}_0/\mathbb{P}_0
=
e^{-\beta \mathcal{H}}/Z (\beta)
\otimes
| 0 \rangle \langle 0 |    
.
$
This state is nothing but the direct product of the normalized Gibbs state and the ancillary $| 0 \rangle$ state.
Since the partition function is defined as the summation of the Boltzmann factors over all the possible states of $n$ qubits,
the $2^n$ terms in it gives $Z(\beta) = \mathcal{O}(2^n),$
leading to $\mathbb{P}_0 = \mathcal{O}(1).$
[See Eq.~(\ref{PITE_at_finite_temperatures:prob_0})]
This fact ensures that, for example,
increasing $n$ for the quantum simulation of a given real molecule to achieve higher resolution of the spatial discretization
will converge to some nonzero $\mathbb{P}_0$ unique to the molecule in the limit of $n \to \infty.$

It is clear that the partition function can be obtained simply from the ratio of the number of outcomes for the success state to the number of measurements,
which immediately gives the free energy via the relation
$e^{-\beta F (\beta)} =  Z (\beta).$
In this sense,
losing in this lottery is not waste of time.
If we did not have the partition function,
the free energy had to be calculated in another way, e.g. the thermodynamic relation $F = E - S/\beta$,
where $E$ is the internal energy and $S$ is the von Neumann entropy.
The calculation of $S$ itself within the framework of PITE seems difficult,
similarly to the case of VITE.\cite{bib:5563}

It is noted here that there exist quantum algorithms for obtaining the complex partition functions for analyses of quantum critical phenomena.\cite{bib:5583, bib:5584, bib:5585}

\section{Applications}
\label{sec:applications}

\subsection{Two-level system}
\label{sec:applications:two_level}

Let us consider a two-level system
for which we assign its ground state and excited state to single-qubit states
$| 0 \rangle$ and $| 1 \rangle$, respectively.
The Hamiltonian of this system is
$
\mathcal{H}
=
\varepsilon_{\mathrm{gs}} | 0 \rangle \langle 0 |
+
\varepsilon_{\mathrm{ex}} | 1 \rangle \langle 1 |
,
$
where $\varepsilon_{\mathrm{gs}}$ and $\varepsilon_{\mathrm{ex}}$
are the energy eigenvalue of ground state and that of excited state, respectively.
We express the input state for the $k$th PITE step $(k = 0, 1, \dots)$ as
$| \psi_k \rangle = \cos \phi_k | 0 \rangle + \sin \phi_k | 1 \rangle$
by using a real mixing angle $\phi_k$.
The operator $\mathcal{M}$ acts in this case as
$
\mathcal{M}
| \psi_k \rangle
=
m_0
(
e^{-\varepsilon_{\mathrm{gs}} \Delta \tau} \cos \phi_k | 0 \rangle
+
e^{-\varepsilon_{\mathrm{ex}} \Delta \tau} \sin \phi_k | 1 \rangle
)
.
$

\subsubsection{For exact circuit}

We first examine the case where
the exact circuit $\mathcal{C}_{\mathrm{PITE}}$ in
Fig.~\ref{circuit:imag_evol_as_part_of_real_evol:evolution_exact_and_st1}(a)
is used for the PITE steps.
By using
\begin{gather}
    \theta_{\lambda}
    \equiv
        \arccos
        \frac{
            m_0
            e^{-\varepsilon_{\lambda} \Delta \tau}
            +
            \sqrt{1 - m_0^2 e^{-2 \varepsilon_{\lambda} \Delta \tau}}
        }{\sqrt{2}}
\end{gather}
$(\lambda = \mathrm{gs, ex}),$
the operator defined in
Eq.~(\ref{imag_evol_as_part_of_real_evol:def_Theta})
for the present system is written as
$
\Theta
= 
\theta_{\mathrm{gs}} | 0 \rangle \langle 0 |
+
\theta_{\mathrm{ex}} | 1 \rangle \langle 1 |
.
$
The unitaries generated by this Hermitian operator are thus
\begin{gather}
    e^{\pm i \kappa \Theta}
    =
        e^{\pm i \kappa \overline{\theta}}
        R_z (\kappa \Delta \theta)
        ,
\end{gather}
where
$
\overline{\theta} 
\equiv
(\theta_{\mathrm{ex}} + \theta_{\mathrm{gs}})/2
$
and
$\Delta \theta \equiv \theta_{\mathrm{ex}} - \theta_{\mathrm{gs}}.$
They constitute the circuit $\mathcal{C}_{\mathrm{PITE}},$
which is specifically depicted in 
Fig.~\ref{fig:circuits_for_two_level}(a).
The normalized success state 
\begin{gather}
    | \psi_{k + 1} \rangle
    =
        \frac{
            | 0 \rangle
            +
            e^{-\Delta \varepsilon \Delta \tau}
            \tan \phi_k | 1 \rangle
        }{\sqrt{1 + e^{-2 \Delta \varepsilon \Delta \tau} \tan^2 \phi_k}}
    \label{imag_time_evol_two_level:state_just_after_measurement}
\end{gather}
is obtained with the probability
\begin{gather}
    p_k 
    =
        m_0^2
        \frac{e^{-2 \varepsilon_{\mathrm{gs}} \Delta \tau} + e^{-2 \varepsilon_{\mathrm{ex}} \Delta \tau} w_k }{1 + w_k }
        ,
    \label{imag_time_evol_two_level:prob_0_using_w1}
\end{gather}
where
$\Delta \varepsilon \equiv \varepsilon_{\mathrm{ex}} - \varepsilon_{\mathrm{gs}}$
is the excitation energy and
$w_k \equiv \tan^2 \phi_k$ is the relative weight of the excited state in the input state.
It is clear from Eq.~(\ref{imag_time_evol_two_level:state_just_after_measurement})
that the mixing angle and the relative weight change as
$
\phi_k 
\longmapsto{}
\phi_{k + 1} = \arctan ( \alpha \tan \phi_k )
$
and
$
w_k 
\longmapsto{}
w_{k + 1} = \alpha^2 w_k,
$
respectively,
where $\alpha \equiv e^{-\Delta \varepsilon \Delta \tau}$ is the decay factor.
We can then obtain $w_k = \alpha^{2 k} w_0$,
indicating that the relative weight decays exponentially as the PITE procedure continues.
The success probability in
Eq.~(\ref{imag_time_evol_two_level:prob_0_using_w1})
is rewritten as a closed form 
$
p_k 
=
m_0^2
(
e^{-2 \varepsilon_{\mathrm{gs}} \Delta \tau}
+
e^{-2 \varepsilon_{\mathrm{ex}} \Delta \tau}
\alpha^{2 k} w_0
)
/(1 + \alpha^{2 k} w_0)
.
$
It is easily proved that $p_{k + 1} > p_k$ for an arbitrary $k$ since $\alpha < 1$.
This means that the success probability at each measurement increases monotonically as the PITE procedure continues and
it reaches the saturated value
$p_{\infty} \equiv m_0^2 e^{-2 \varepsilon_{\mathrm{gs}} \Delta \tau}$,
independent of the initial relative weight.

\subsubsection{For approximate circuit}

\begin{figure}
\begin{center}
\includegraphics[width=8.5cm]{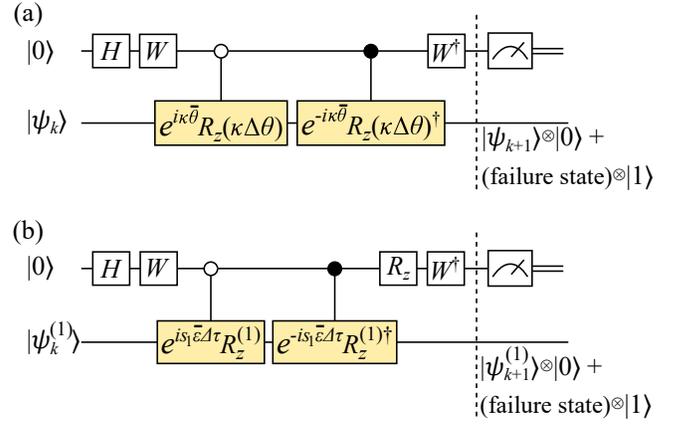}
\end{center}
\caption{
(a) Exact $\mathcal{C}_{\mathrm{PITE}}$ and
(b) approximate $\mathcal{C}_{\mathrm{PITE}}^{(1)}$
PITE circuits for the two-level system,
as special cases of those in 
Fig.~\ref{circuit:imag_evol_as_part_of_real_evol:evolution_exact_and_st1}.
The input register in each of them consists of a single qubit.
$R_z^{(1)} \equiv R_z (-s_1 \Delta \varepsilon \Delta \tau)$
is used in this figure.
The input state $| \psi_k \rangle$ to $\mathcal{C}_{\mathrm{PITE}}$ for the $k$th step undergoes the gate operations to form the state shown near the dashed line, immediately before the measurement.
This state is entangled with the ancilla and contains the success state $| \psi_{k + 1} \rangle$ for the next step.
It is similarly the case with $| \psi_k^{(1)} \rangle$ input to $\mathcal{C}_{\mathrm{PITE}}^{(1)}.$
}
\label{fig:circuits_for_two_level}
\end{figure}

We next examine the case where the approximate circuit
$\mathcal{C}_{\mathrm{PITE}}^{(1)}$ in
Fig.~\ref{circuit:imag_evol_as_part_of_real_evol:evolution_exact_and_st1}(b) is used.
The RTE operator for the present system
\begin{gather}
    U_{\mathrm{RTE}} (\Delta t)
    =
        e^{-i \overline{\varepsilon} \Delta t}
        R_z (-\Delta \varepsilon \Delta t)
        ,
\end{gather}
where
$
\overline{\varepsilon}
\equiv
(\varepsilon_{\mathrm{ex}} + \varepsilon_{\mathrm{gs}} )/2,
$
constitutes the circuit, as depicted in
Fig.~\ref{fig:circuits_for_two_level}(b).
The success probability at the $k$th measurement is calculated as
\begin{gather}
    p_k^{(1)}
    =
        \frac{1}{2}
        +
        \frac{
            \sin
            (2 \theta_0 - 2 \varepsilon_{\mathrm{gs}} s_1 \Delta \tau)
            +
            \sin
            (2 \theta_0 - 2 \varepsilon_{\mathrm{ex}} s_1 \Delta \tau)
            w_k^{(1)}
        }{2 (1 + w_k^{(1)})}
        ,
    \label{imag_time_evol_two_level:prob_0_using_w1_tau_1}
\end{gather}
where $w_k^{(1)}$ is the relative weight of excited state in this case.
The normalized success state is
\begin{gather}
| \psi_{k + 1}^{(1)} \rangle 
=
    \frac{
        \gamma_{\mathrm{gs}}
        \cos \phi_k
        | 0 \rangle
        +
        \gamma_{\mathrm{ex}}
        \sin \phi_k
        | 1 \rangle
    }{\sqrt{2 p_k^{(1)}}}
    ,
\end{gather}
where 
$
\gamma_\lambda
\equiv
\cos
(\theta_0 - \varepsilon_\lambda s_1 \Delta \tau)
+
\sin
(\theta_0 - \varepsilon_\lambda s_1 \Delta \tau)
\ 
(\lambda = \mathrm{gs}, \mathrm{ex})
.
$
By introducing the decay factor
$\alpha' \equiv \gamma_{\mathrm{ex}}/\gamma_{\mathrm{gs}}$
for the approximate circuit,
we obtain
$w_k^{(1)} = \alpha'^{2 k} w_0$ for the exponential decay of the relative weight
similarly to the case of exact circuit.
The number of steps necessary for the relative weight to be $\delta$ is thus estimated to be
$
n_{\mathrm{steps}} (\delta) 
\approx
-(1 + \tan \theta_0)
\ln (\delta/w_{0} )/(2 s_1 \Delta \varepsilon \Delta \tau)
.
$

\begin{figure}
\begin{center}
\includegraphics[width=6cm]{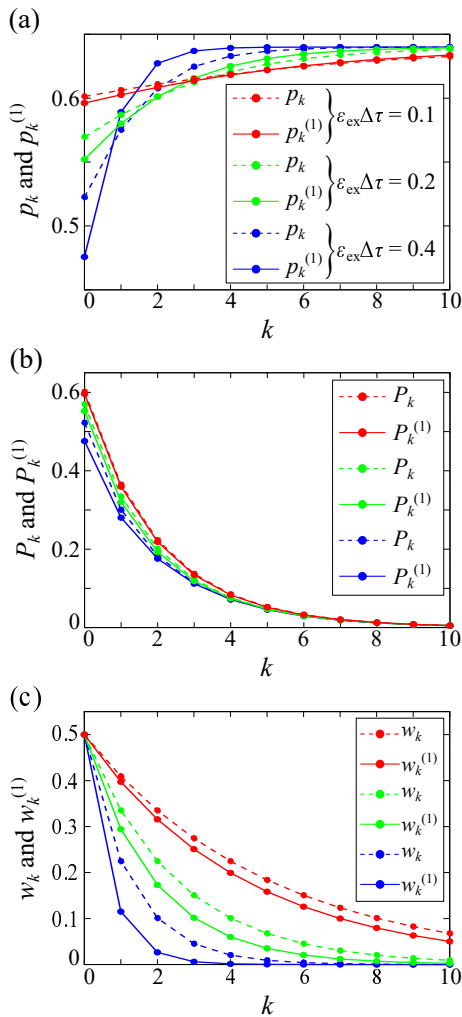}
\end{center}
\caption{
For $\varepsilon_{\mathrm{gs}} = 0$, $m_0 = 0.8$, and some values of $\Delta \tau$,
the success probabilities for the exact $\mathcal{C}_{\mathrm{PITE}}$ and approximate $\mathcal{C}_{\mathrm{PITE}}^{(1)}$ circuits at each step $k$ are plotted in (a).
The probability $P_k$ that all the measurements in the exact circuit by the end of $k$th step are successful is plotted in (b).
The similarly defined probability $P_k^{(1)}$ for the approximate circuit is also plotted.
(c) is the plot of the relative weights of excited states.
}
\label{fig:two_level}
\end{figure}

The success probabilities $p_k$ for the exact circuit and
$p_k^{(1)}$ for the approximate circuit at each step as functions of $k$
for $\varepsilon_{\mathrm{gs}} = 0$ and $m_0 = 0.8$ are plotted
in Fig.~\ref{fig:two_level}(a).
Although those probabilities at $k = 0$ are smaller for a larger $\Delta \tau$ as expected,
the order is reversed beyond for $k > 3$ and they increase monotonically toward the saturated value.
Figure \ref{fig:two_level}(b) is the plot of
the probability $P_k \equiv \prod_{k' = 0}^k p_{k'}$ that all the measurements in the exact circuit by the end of $k$th step are successful.
The probability $P_k^{(1)}$ for the approximate circuit is also defined similarly.
We can see that the differences between the probabilities at each $k$ in terms of the circuits and $\Delta \tau$ become smaller as the steps continue.
Figure \ref{fig:two_level}(c) is the plot of
the relative weights $w_k$ and $w_k^{(1)}$. 
For both of the circuits, the weights decay more rapidly for a larger $\Delta \tau$,
as explained above.
The approximate circuit exhibits the smaller weight than the exact circuit does.
Apart from this favorable behavior,
which might be accidental in the present case,
the asymptotic behavior found for this system can be qualitatively true for generic systems.
It is because any system whose state is sufficiently close to the ground state
can be treated approximately as a two-level system if there is no degeneracy.

\subsection{Wave functions for an asymmetric double-well potential}
\label{sec:applications:asymmetric}

We consider here a quantum mechanical particle confined to an asymmetric double-well potential in one-dimensional space by using the approach described in
Sect.~\ref{sec:methods:PITE_for_real_space}.

The potential shape is drawn as the black curve in Fig.~\ref{fig:double_well}(a),
whose expression is provided in Appendix~\ref{sec:appendix:asymmetric_pot}.
We encoded the wave function of the particle of unit mass at the discretized points by using six qubits.
The energy eigenstates of this system, also shown in the figure, are obtained by numerically diagonalizing the Hamiltonian matrix of dimension $2^6 = 64$.

Given the double-well structure,
it is reasonable to adopt an initial state for the PITE steps having two peaks at the locations of minima.
By using the normalized Gaussian wave function
$| g (x_{\mathrm{c}}, \sigma) \rangle$
centered at $x_{\mathrm{c}}$ with width $\sigma$,
we defined the initial state
$
| \psi_{\mathrm{init}}^{(\mathrm{d})} \rangle
\propto
| g (L/2 + d/2, d/3) \rangle
+
| g (L/2 - d/2, d/3) \rangle/2
$
with the length $L$ of simulation cell and the distance $d$ between the potential minima.
We simulated the PITE procedure using $m_0 \equiv 0.9$ and $\Delta \tau \equiv 0.1$.
Some of the success states $| \psi_k \rangle$ during the steps are plotted in
Fig.~\ref{fig:double_well}(b),
where the initial state converges correctly to the ground state.
The probability $p_k$ for obtaining the success state at each step is plotted in Fig.~\ref{fig:double_well}(c),
where it converges to the saturated value close to $m_0^2$.
The weight of each energy eigenstate in the input state at each step is plotted in 
the left panel of Fig.~\ref{fig:double_well}(d).
Since the weight of ground state is rather high already in the initial state,
the rapid convergence is achieved.

To see the effects of initial state on the success states during the steps,
we also defined the initial state
$
| \psi_{\mathrm{init}}^{(\mathrm{s})} \rangle
=
| g (L/2 + d/2, d/3) \rangle
$
having only a single peak at the lower minimum
and simulated the PITE procedure.
The weight of each energy eigenstate is plotted in 
the right panel of Fig.~\ref{fig:double_well}(d).
On the contrary to the case of
$| \psi_{\mathrm{init}}^{(\mathrm{d})} \rangle$,
the convergence is rather slow.
This result and that for a harmonic potential
(see Appendix \ref{sec:applications:parabolic})
tell us that the rapid convergence to the ground state in a PITE calculation requires a good initial guess,
which might be also the case for VITE and QITE calculations. 
When a molecular system is given and we try to decide an initial guess for PITE steps,
a typical way for finding a good guess is to perform a cheap classical calculation such as that based on the Hartree--Fock theory\cite{Helgaker} or the density functional theory.~\cite{bib:76, bib:77}
The following two points should, however, be kept in mind.
First, such cheap calculations may not give a good initial guess for a molecule where electronic correlation is strong.
Second, the state preparation for an adopted initial guess via gate operations become more complicated in general as we treat a larger molecule and/or the spatial resolution for the guess is higher.
The second point will force us to find compromise between the resolution and the operation number in the state preparation.

\begin{figure*}
\begin{center}
\includegraphics[width=15cm]{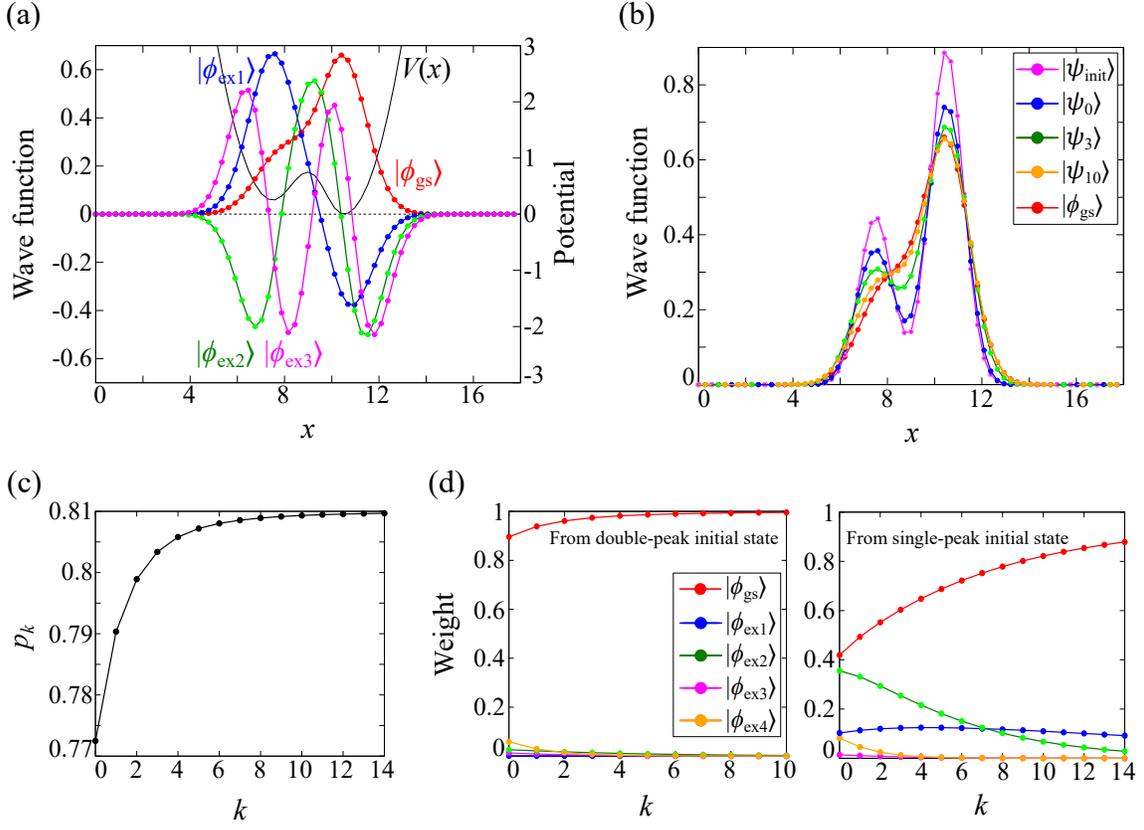}
\end{center}
\caption{
(a)
Circles represent the exact ground state $| \phi_{\mathrm{gs}} \rangle$
and the three lowest excited states
$| \phi_{\mathrm{ex} \mu} \rangle \ (\mu = 1, 2, 3)$
at the discrete points encoded by six qubits
for the particle in the asymmetric double-well potential $V (x)$,
shown as the black curve.
(b)
Red and purple circles represent $| \phi_{\mathrm{gs}} \rangle$ and
$| \psi_{\mathrm{init}}^{(\mathrm{d})} \rangle,$ respectively.
The other circles represent the success state
$| \psi_k \rangle \ (k = 0, 1, \dots)$
obtained immediately after the measurement at the PITE step $k$. 
(c)
Probability for obtaining the success state at each PITE step.
(d)
Left panel shows the weight of each energy eigenstate in the input state at each step
for the double-peak initial state
$| \psi_{\mathrm{init}}^{(\mathrm{d})} \rangle$,
while the right panel shows that for the single-peak initial state
$| \psi_{\mathrm{init}}^{(\mathrm{s})} \rangle$.
}
\label{fig:double_well}
\end{figure*}

\section{Summary}
\label{sec:conclusions}

In summary, we have proposed a generic approach of PITE which constructs the circuit from the RTE gates as black boxes within the first-order of imaginary-time step. 
We discussed the overall PITE procedure of quantum chemistry calculations
and the scaling of computational cost to propose the FQE framework.
As demonstrated in the simple examples,
the PITE approach suffers from the inherent drawback,
i.e., the probabilistic nature.
If we overcome it to some extent with the aid of the QAA techniques,
we will be able to concentrate on the development of efficient unitary algorithms for real-time dynamics.
The examinations on the quality of existing and new QAA techniques adapted to PITE will thus be important in the future in order for this nonvariational approach to be a standard for practical quantum computation.

\begin{acknowledgments}

This work was supported by MEXT as "Program for Promoting Researches on the
Supercomputer Fugaku" (JPMXP1020200205) and JSPS KAKENHI as "Grant-in-Aid for Scientific Research(A)" Grant Number 21H04553.
The computation in this work has been done using (supercomputer Fugaku provided by the RIKEN Center for Computational Science/Supercomputer Center at the Institute for Solid State Physics in the University of Tokyo).

\end{acknowledgments}


\appendix

\section{Change in an input state undergoing exact PITE circuit}
\label{sec:state_change_during_exact_circuit}

We describe here the change in an input many-qubit state $| \psi \rangle$ via the gate operations in the exact PITE circuit $\mathcal{C}_{\mathrm{PITE}}$ in 
Fig.~\ref{circuit:imag_evol_as_part_of_real_evol:evolution_exact_and_st1}(a).

Before examining the circuit,
we provide an additional explanation of the definition of $\Theta.$
Since the inverse trigonometric function arccos for an operator is defined via Taylor series as usual and its value is within the range $(0, \pi),$
$\Theta$ defined in
Eq.~(\ref{imag_evol_as_part_of_real_evol:def_Theta})
falls into this range.
We have then
$
\sin \Theta
=
|(\mathcal{M} - \sqrt{1-\mathcal{M}^2})/\sqrt{2}|
,
$
where $\mathcal{M}$ is understood as each of its eigenvalues,
and the issue is the sign of the RHS.
Considering a case for $\Delta \tau = 0,$
the sign of $\mathcal{M} - \sqrt{1 - \mathcal{M}^2}$
changes across $m_0 = 1/\sqrt{2}.$
We therefore defined $\kappa$ as in the main text and found the expression
$
\sin \Theta
=
\kappa (\mathcal{M}-\sqrt{1-\mathcal{M}^2})/\sqrt{2}
$
to give the correct (positive) $\sin \Theta.$
In addition, by appropriately redefining $m_0$,
we can set the origin of energy in the ITE operator sufficiently close to the lowest energy eigenvalue,
so that the Taylor expansion in $\mathcal{H} \Delta \tau$ is justified and $\sqrt{1 - \mathcal{M}^2}$ is real.

The composite state of the input register and the initialized ancilla changes via the gate operations as
\begin{gather}
    | \psi \rangle
    \otimes
    | 0 \rangle
    \xmapsto{H, W}
        | \psi \rangle
        \otimes
        \left(
            \frac{1 - i}{2} | 0 \rangle
            +
            \frac{1 + i}{2} | 1 \rangle
        \right)
    \nonumber \\
    \xmapsto{e^{i \kappa \Theta}, e^{-i \kappa \Theta}}
            e^{i \kappa \Theta}
            | \psi \rangle
            \otimes
            \frac{1 - i}{2} | 0 \rangle
            +
            e^{-i \kappa \Theta}
            | \psi \rangle
            \otimes
            \frac{1 + i}{2} | 1 \rangle
    \nonumber \\
    \xmapsto{W^\dagger}
        \frac{(1 - i) e^{i \kappa \Theta} + (1 + i) e^{-i \kappa \Theta}}
        {2 \sqrt{2}}
        | \psi \rangle
        \otimes
        | 0 \rangle
    \nonumber \\
        +
        \frac{(1 + i) e^{i \kappa \Theta} + (1 - i) e^{-i \kappa \Theta}}
        {2 \sqrt{2}}
        | \psi \rangle
        \otimes
        | 1 \rangle
    \nonumber \\
    =
        \frac{\cos \Theta + \kappa \sin \Theta}{\sqrt{2}}
        | \psi \rangle
        \otimes
        | 0 \rangle
        +
        \frac{\cos \Theta - \kappa \sin \Theta}{\sqrt{2}}
        | \psi \rangle
        \otimes
        | 1 \rangle
        ,
\end{gather}
where the RHS of the equality gives the resultant state in 
Eq.~(\ref{imag_evol_as_part_of_real_evol:whole_state_before_measurement}).

\section{Probabilistic evolution for a non-Hermitian generator}
\label{sec:prob_evol_for_non_Hermitian}

We describe here the algorithm for probabilistic evolution of a quantum state
$| \psi (t) \rangle$
whose dynamics is governed by a differential equation of the form
$d | \psi (t) \rangle/dt = \mathcal{L} | \psi (t) \rangle$.
The formal solution for a time step $\Delta t$ is 
$| \psi (t + \Delta t) \rangle = e^{\mathcal{L} \Delta t} | \psi (t) \rangle.$
If the generator $\mathcal{L}$ of the evolution is anti-Hermitian,
the dynamics is unitary and thus easy to implement without any measurement.
If the generator is Hermitian, $\mathcal{C}_{\mathrm{PITE}}$ circuit or its approximate version described above is applicable with an ancillary qubit.
Therefore we assume it to be non-Hermitian,
for which we demonstrate its probabilistic evolution can be implemented by using two ancillary qubits.

\begin{figure*}
\begin{center}
\includegraphics[width=11cm]{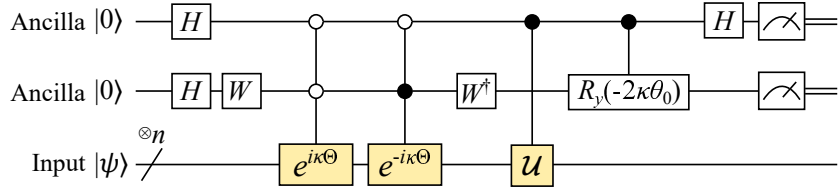}
\end{center}
\caption{
$(n + 2)$-qubit circuit $\mathcal{C}_{\mathcal{L}}$ for the real-time dynamics generated by a non-Hermitian operator $\mathcal{L}.$
The yellow boxes represent the gates involving $\mathcal{L}.$
}
\label{circuit:time_evol_from_non_hermitian:evolution}
\end{figure*}

Since $\mathcal{L} + \mathcal{L}^\dagger$ is Hermitian,
we can define a nonunitary Hermitian operator
\begin{gather}
    \mathcal{M}
    \equiv
        m_0
        e^{(\mathcal{L} + \mathcal{L}^\dagger) \Delta t}
        ,
    \label{time_evol_from_non_hermitian:def_M}
\end{gather}
where $m_0$ is a constant similar to the case for PITE.
In addition,
by using the fact that $\mathcal{L} - \mathcal{L}^\dagger$ is anti-Hermitian,
we define a non-Hermitian unitary operator
\begin{gather}
    \mathcal{U}
    \equiv
        e^{ (\mathcal{L} - \mathcal{L}^\dagger) \Delta t}
        .
\end{gather}
We define the operator $\Theta$ and the constants $\kappa$ and $\theta_0$
similarly to the case of PITE.
With them, we introduce two ancillary qubits to construct the circuit 
$\mathcal{C}_{\mathcal{L}}$ for the $(n + 2)$-qubit system,
as shown in Fig.~\ref{circuit:time_evol_from_non_hermitian:evolution}.
The composite system undergoes the unitary operations as
\begin{gather}
    | \psi \rangle
    \otimes
        | 0 \rangle
        \otimes
        | 0 \rangle
    \nonumber \\
    \longmapsto{}
        \frac{1}{2}
            (\mathcal{M} + m_0 \mathcal{U} )
            | \psi \rangle
            \otimes
            | 0 \rangle
            \otimes
            | 0 \rangle
        +
        (\mathrm{other \ states})
        ,
    \label{time_evol_from_non_hermitian:whole_state_before_measurement}
\end{gather}
which is the state immediately before the measurement on the ancillae.
The $n$-qubit state coupled to the ancillary state
$| 0 \rangle \otimes | 0 \rangle$
in Eq.~(\ref{time_evol_from_non_hermitian:whole_state_before_measurement})
is
\begin{gather}
    \frac{1}{2}
    (\mathcal{M} + m_0 \mathcal{U})
    | \psi \rangle
    =
        m_0
        \left(
            1 + \mathcal{L} \Delta t
        \right)
        | \psi \rangle
        +
        \mathcal{O} (\Delta t^2)
    ,
\end{gather}
which coincides with the correct evolved state $e^{\mathcal{L} \Delta t} | \psi \rangle$
within the first order of $\Delta t$.
If the measurement outcome is $| 0 \rangle \otimes | 0 \rangle$,
the success state $e^{\mathcal{L} \Delta t} | \psi \rangle$
with a normalization constant will be obtained.
It is easily understood that
the success state via the steps will eventually fall onto the subspace corresponding the Jordan block having the lowest real part of eigenvalue among the blocks.

It is noted here that
there exists a quantum algorithm for solving an inhomogeneous linear equation
proposed by Xin et al.\cite{bib:5052}
They expand the formal solution in time up to a finite order,
for which the expansion coefficients are obtained probabilistically by introducing ancillary bits.

\section{Kinetic propagator for one-dimensional dynamics}
\label{sec:kinetic_propagator}

\begin{figure}
\begin{center}
\includegraphics[width=7cm]{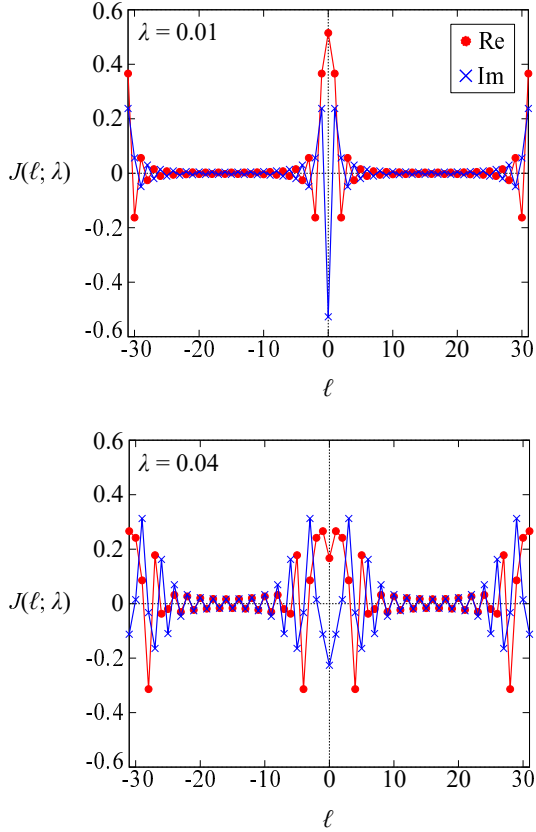}
\end{center}
\caption{
Kinetic propagator $J (\ell; \lambda)$ for $n = 5 \ (N = 32)$.
The upper and lower panels are for $\lambda = 0.01$ and $0.04$, respectively.
}
\label{fig:kin_propagator}
\end{figure}

From eqs.~(\ref{QITE_as_a_part_of_RITE:def_mom_eigenstate}) and 
(\ref{imag_evol_as_part_of_real_evol:CQFT_on_kinetic_rotation}),
the action of kinetic-evolution operator on the position eigenstate $| k \rangle_n \ (k = 0, \dots, N - 1)$ is explicitly calculated as
\begin{gather}
    e^{-i \hat{T} \Delta t}
    | k  \rangle_n
    =
        \sum_{s = 0}^{N - 1}
            e^{-i E_{s} \Delta t}
            | p^{(\widetilde{s})} \rangle
            \langle p^{(\widetilde{s})}
            | k  \rangle_n
    \nonumber \\
    =
        \frac{1}{N}
        \sum_{s = 0}^{N - 1}
            \exp
            \left(
                -i
                \widetilde{s}^2
                \frac{(\Delta p)^2}{2 m}
                \Delta t
            \right)
    \cdot
    \nonumber \\
    \cdot
        \sum_{k' = 0}^{N - 1}
            e^{-i \pi (k' - k  )}
            \exp \frac{2 \pi i s (k' - k  )}{N}
            | k' \rangle_n
    \nonumber \\
    =
        \sum_{k' = 0}^{N - 1}
            J
            \left(
                k' - k  ;
                \frac{(\Delta p)^2}{2 m}
                \Delta t
            \right)
            | k' \rangle_n
    ,
\end{gather}
where we have defined the kinetic propagator
\begin{gather}
    J (\ell; \lambda)
    \equiv
        \frac{1}{N}
        e^{-i \pi \ell}
        \sum_{s = 0}^{N - 1}
            \exp
            \left(
                -i
                \lambda
                \widetilde{s}^2
                +
                \frac{2 \pi i \ell s}{N}
            \right)
\end{gather}
for $\ell = -N + 1, -N + 2, \dots, N - 1.$
This satisfies clearly the symmetry $J (\ell; \lambda) = J (-\ell; \lambda)$
and the probability conservation 
$\sum_{\ell = 0}^{N - 1} J (\ell; \lambda) = 1.$
The propagator for five qubits is plotted in Fig.~\ref{fig:kin_propagator} as an example.
We can understand easily that
the periodic boundary condition for kinetic-evolution has entered naturally our formulation
and that the evolution operator affects the amplitude of wave function mainly around a point at which the particle is located.

The matrix element of kinetic-energy operator is similarly calculated as
\begin{gather}
    \langle k |_n \hat{T} | k' \rangle_n
    =
        \frac{e^{-i \pi (k - k')}}{N}
        \sum_{s = 0}^{N - 1}
            E_s
            \exp \frac{2 \pi i (k - k') s}{N}
    ,
\end{gather}
which are used in the numerical diagonalization of Hamiltonians in 
Sect \ref{sec:applications:asymmetric}
and
Appendix \ref{sec:applications:parabolic}.

\section{Shape of double-well potential}
\label{sec:appendix:asymmetric_pot}

We defined the potential for the length $L \equiv 18$ of simulation cell as
\begin{gather}
    V (x)
    =
    \begin{cases}
        ( x - \frac{L}{2} + \frac{d}{2} )^2/2
        + \Delta 
        &
        x \leqq \frac{L - d}{2}
        \\
        \frac{V_0}{2} (1 + \cos [ \frac{2 \pi}{d} (x - \frac{L}{2})])
        + \Delta
        &
        \frac{L - d}{2} < x \leqq \frac{L}{2}
        \\
        \frac{V_0 + \Delta}{2}
        (1 + \cos [ \frac{2 \pi}{d} (x - \frac{L}{2})])
        &
        \frac{L}{2} < x \leqq \frac{L + d}{2}
        \\
        (x - \frac{L}{2} - \frac{d}{2})^2/2
        &
        x > \frac{L + d}{2}
        \\
    \end{cases}
    ,
\end{gather}
where $d \equiv 3$ is the distance between the minima and
$\Delta \equiv 0.25$ is the height of higher minimum measured from the lower minimum.
$V_0 \equiv 0.5$ is the strength of barrier between the minima.

\section{Wave functions for a parabolic potential}
\label{sec:applications:parabolic}

We consider here a quantum mechanical particle confined to a parabolic potential in one-dimensional space by using the approach described in
Sect.~\ref{sec:methods:PITE_for_real_space}.

\begin{figure*}
\begin{center}
\includegraphics[width=15cm]{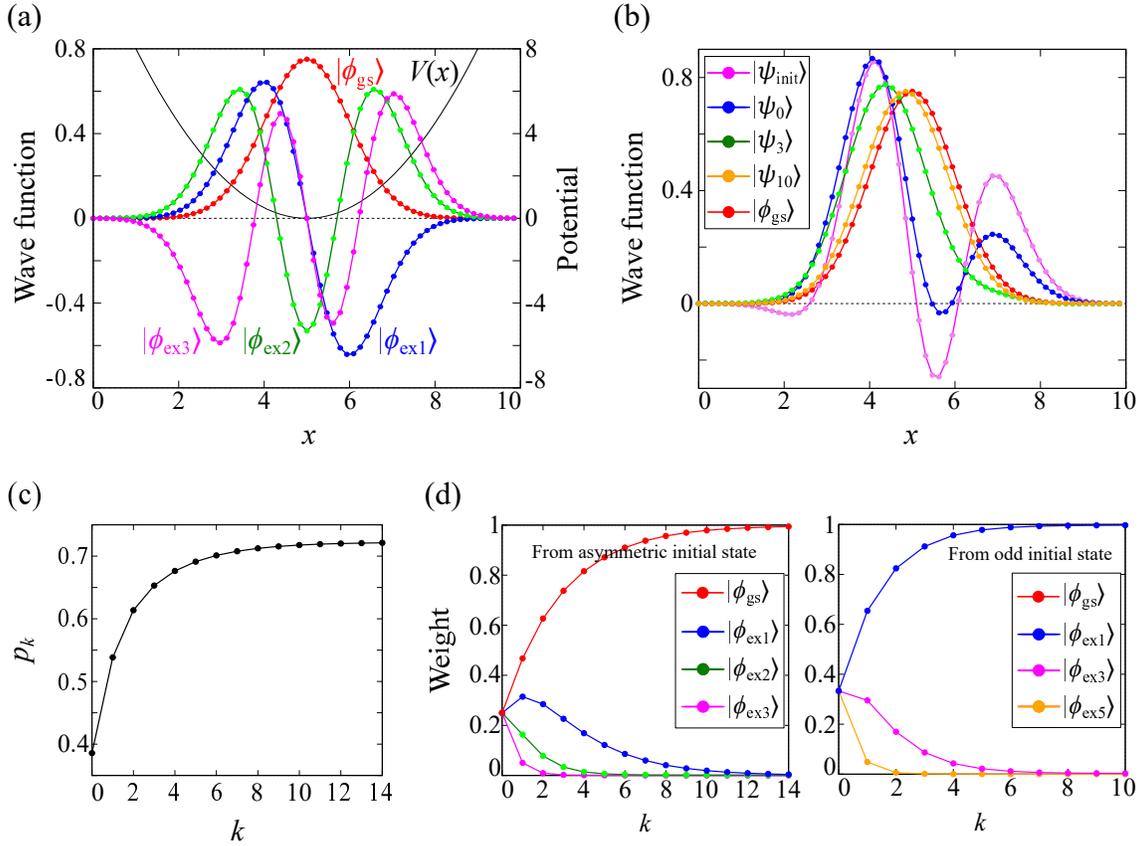}
\end{center}
\caption{
(a)
Circles represent the exact ground state $| \phi_{\mathrm{gs}} \rangle$
and the three lowest excited states
$| \phi_{\mathrm{ex} \mu} \rangle \ (\mu = 1, 2, 3)$
at the discrete points encoded by six qubits
for the particle in the harmonic potential $V (x)$,
shown as the black curve.
(b)
Red and purple circles represent $| \phi_{\mathrm{gs}} \rangle$ and
$| \psi_{\mathrm{init}} \rangle,$ respectively.
The other circles represent the success state
$| \psi_k \rangle \ (k = 0, 1, \dots)$
obtained immediately after the measurement at the PITE step $k$. 
(c)
Probability for obtaining the success state at each PITE step.
(d)
Left panel shows the weight of each energy eigenstate in the input state at each step
for the asymmetric initial state
$| \psi_{\mathrm{init}} \rangle$,
while the right panel shows that for the odd initial state $| \psi_{\mathrm{init}}^{(\mathrm{odd})} \rangle$ with $\Delta \tau = 0.1$.
}
\label{fig:harmonic}
\end{figure*}

We defined the potential for the length $L \equiv 10$ of simulation cell as
$V (x) = m \omega^2 (x - L/2)^2/2$ with $\omega \equiv 1$.
We encoded the wave function of the particle of unit mass at the discretized points by using six qubits.
The energy eigenstates of this system are obtained by numerically diagonalizing the Hamiltonian matrix of dimension $2^6 = 64$,
as shown in Fig.~\ref{fig:harmonic}(a).

We simulated the PITE procedure for this system by adopting
the circuit $\mathcal{C}_{\mathrm{PITE}}^{(\mathrm{ST1})}$ in
Fig.~\ref{fig:pite_circuit_for_1dim_and_electrons}(a).
We set $m_0 \equiv 0.85$ and the imaginary-time step $\Delta \tau \equiv 0.15$ for the operator $\mathcal{M}$ with the Hamiltonian shifted according to the ground state energy.
We prepared the superposition of the ground state and the three lowest excited states as the initial state
$
| \psi_{\mathrm{init}} \rangle
=
(
| \phi_{\mathrm{gs}} \rangle
+
| \phi_{\mathrm{ex1}} \rangle
+
| \phi_{\mathrm{ex2}} \rangle
+
| \phi_{\mathrm{ex3}} \rangle
)/2
$
for the PITE steps.
Some of the success states $| \psi_k \rangle$ during the steps are plotted in
Fig.~\ref{fig:harmonic}(b),
where the initial double-peak shape of the wave function has disappeared at the step $k = 3$ and the wave function at $k = 10$ already has a large overlap with the ground state.
The probability $p_k$ for obtaining the success state at each step is plotted in Fig.~\ref{fig:harmonic}(c),
where it converges to the saturated value close to $m_0^2$.
The weight of each energy eigenstate in the input state at each step is plotted in 
the left panel of Fig.~\ref{fig:harmonic}(d),
where the weight of ground state increases monotonically
while those of the second and third excited states decay rapidly.

We also prepared the superposition of the excited states with odd parity as the initial state
$
| \psi_{\mathrm{init}}^{(\mathrm{odd})} \rangle
=
(
| \phi_{\mathrm{ex1}} \rangle
+
| \phi_{\mathrm{ex3}} \rangle
+
| \phi_{\mathrm{ex5}} \rangle
)/\sqrt{3}
$
and simulated the PITE procedure with $\Delta \tau \equiv 0.1$.
The weight of each energy eigenstate during the steps are plotted in
the right panel of Fig.~\ref{fig:harmonic}(d),
where only the first excited state survives the steps and the other two states decay to zero.
This result indicates that the odd parity in the initial state was preserved correctly during the steps and the lowest-energy state $|\phi_{\mathrm{ex1}} \rangle$ within the odd-parity subspace was obtained since the initial state had an overlap with it.

\bibliography{ref}

\end{document}